\documentstyle[aps,amssymb,preprint,graphics]{revtex}

\begin{document}

\title{Iterative Approach to Gravitational Lensing Theory } \author{Thomas P.  Kling, Ezra 
T.
Newman, and Alejandro Perez \\ Dept.  of Physics and Astronomy,\\ University of Pittsburgh,
Pgh, PA, 15260, USA} \date{\today} \maketitle

\begin{abstract} We develop an iterative approach to gravitational lensing theory based on
approximate solutions of the null geodesic equations.  The approach can be employed in any
space-time which is ``close'' to a space-time in which the null geodesic equations can be
completely integrated, such as Minkowski space-time, Robertson-Walker cosmologies, or
Schwarzschild-Kerr geometries.  To illustrate the method, we construct the iterative
gravitational lens equations and time of arrival equation for a single Schwarzschild lens.
This example motivates a discussion of the relationship between the iterative approach, the
standard thin lens formulation, and an exact formulation of gravitational lensing.
\end{abstract}

\section{Introduction}

In several recent papers \cite{FN,EFN}, an exact approach to gravitational lensing has been
developed in which a parametric represention of the past light-cone of an observer is viewed 
as the fundamental gravitational lens equations.  In this
approach, one would solve, {\it in principle,} the null geodesic equations of general
relativity in an arbitrary space-time and construct the past light-cone of an observer.  The
lens and time of arrival equations then follow from a particular parametric representation 
of
the past light cone.  The exact approach can be implemented in only a few space-times
possessing a high degree of symmetry, such as Minkowski space, Robertson-Walker cosmologies,
or Schwarzschild/Kerr geometries.  We will refer to space-times in which the null geodesic
equations can be integrated exactly as integrable geometries.

The exact approach stands in contrast to the generally accepted thin lens approximation in
which one does not attempt to find the exact null geodesics.  In this approximation, one
introduces sharp bending at isolated points near the lens along a geodesic trajectory of a
background space-time -- usually either Minkowski space or a cosmological model.  The thin
lens approximation has proved extremely valuable to practicing astrophysicists as a
relatively accurate method which is easy to implement \cite{Ehlers}.

Our goal in this paper is to propose a middle ground in which one systematically finds
approximate null geodesics in geometries ``close'' to integrable geometries and uses these
approximate geodesics as the basis for gravitational lensing calculations.  These geometries
need not correspond to any solution of the Einstein equations; our method should apply to 
any
approximate metric which can be considered as a perturbation of an exact integrable 
geometry.
A key component of this project is a perturbation method known as the ``variation of
constants,' which can be found in several classical mechanics textbooks
\cite{Goldstein,Corben}.  This method is based on techniques from Hamiltonian mechanics and
is described in Section~II.  Though it is often used for perturbations of periodic or
quasi-periodic orbits, it is suited to finding approximations to arbitrary trajectories,
regardless of their type.

In Section~III, the method is specialized to the problem of finding approximate null
geodesics in space-times close to integrable ones, and turning these geodesics into lens
equations.  As an example of the method, we apply the iterative approach to a single
Schwarzschild lens in Section~IV.  In a final section, we discuss the relationship between
the iterative, thin lens, and exact formulations of lensing theory.

\section{Finding Approximate Trajectories}

In this section, we describe a method to obtain approximate trajectories for a perturbed
Hamiltonian.  The discussion largely follows a similar discussion in Goldstein's {\it
Classical Mechanics} \cite{Goldstein}.

Throughout this paper, we refer to an unperturbed Hamiltonian, $H_{o}$, and a perturbed
Hamiltonian,

\begin{equation} H(x^{a},p_{a})=H_{o}(x^{a},p_{a})+\Delta H(x^{a},p_{a}).  \end{equation}

\noindent In addition, we will make no distinction between the time coordinate, $t$, and the
spatial coordinates, $x^{i}$.  The base space coordinates, $x^{a} $, are thus
$x^{a}=(t,x^{i})$, and the momentum coordinates, $p_{a}$, include $p_{0}$, which is
canonically conjugate to $t$.  A parameter, $\lambda$, will be an affine parameterization of
the trajectory, and is a linear function of $t$ if the Hamiltonian does not depend on time.
A dot derivative will always refer to a total derivative with respect to $\lambda$, while
partial $\lambda$ derivatives will always be explicitly written.

The method begins by assuming that one can find a complete solution,
$F_{o}(x^{a},P_{a},\lambda )$, to the Hamilton-Jacobi equation for the unperturbed
Hamiltonian,

\begin{equation} H_{o}(x^{a},\frac{\partial F_{o}}{\partial x^{a}})+\frac{\partial
F_{o}}{\partial \lambda } = 0, \end{equation}

\noindent with $n$ constants of integration, $P_{a}$.  When

\begin{equation} \det \frac{\partial^2 F_o}{\partial x^a \, \partial P_b} \ne 0,
\end{equation}

\noindent $F_{o}(x^{a},P_{a},\lambda)$ is the generator of a (parameter dependent) canonical
transformation, $(x^{a},p_{a}) \Rightarrow (X^{a},P_{a})$.  The transformation is defined by

\begin{eqnarray} X^{a} &=&\frac{\partial F_{o}}{\partial P_{a}}\equiv
X^{a}(x^{a},P_{a},\lambda ) \label{X} \\ p_{a} &=&\frac{\partial F_{o}}{\partial 
x^{a}}\equiv
p_{a}(x^{a},P_{a},\lambda ), \label{p} \end{eqnarray}

\noindent with the inversion of Eq.(\ref{X}) being

\begin{equation} x^a = \chi^a(X^a, P_a, \lambda).  \label{inver} \end{equation}

\noindent When the inversion is substituted into Eq.(\ref{p}), one has the canonical
coordinate transformation in the form

\begin{eqnarray} x^{a} &=&\chi^{a}(X^{a},P_{a},\lambda ) \nonumber \\ p_{a} &=& p_{a}\left(
x^{a}(X^{a},P_{a},\lambda ),P_{a},\lambda )=\pi_{a}(X^{a},P_{a},\lambda \right) .
\label{canntrans} \end{eqnarray}

\noindent It is important to note that Eq.(\ref{canntrans}) represents the solution to the
equations of motion of the unperturbed Hamiltonian in terms of $2n$ constants, $(X^a, P_a)$,
which represent the initial location and momentum, and the affine parameter, $\lambda$.

Since the property of being a canonical transformation is independent of the particular form
of the Hamiltonian, we may apply the canonical transformation generated by the solution to
the unperturbed Hamilton-Jacobi equation to the perturbed Hamiltonian.  Under any canonical
transformation generated by the complete solution, $F_{o}(x^a, P_a, \lambda)$, the
Hamiltonian becomes

\begin{equation} H^{\prime }(X^{a},P_{a},\lambda )=H \left( x^{a}(X^{a},P_{a},\lambda )
,p_{a}(X^{a},P_{a},\lambda ) \right)+\frac{\partial F_{o}}{\partial \lambda } .
\end{equation}

\noindent Using $H=H_{o}+\Delta H$, we have

\begin{equation} H^{\prime}(X^a, P_a, \lambda) = \Delta H \left( x^a (X^a, P_a, \lambda), 
p_a
(X^a, P_a, \lambda) \right) , \end{equation}

\noindent because $F_{o}(x^{a},P_{a},\lambda )$ was chosen to satisfy the Hamilton-Jacobi
equation of the unperturbed Hamiltonian.

Hamilton's equations of motion for $X^a$ and $P_a$ are

\begin{eqnarray} {\dot X}^a (X^a, P_a, \lambda) &=& ~\frac{\partial H^{\prime}}{\partial 
P_a}
(X^a, P_a, \lambda) \nonumber \\ {\dot P}_a (X^a, P_a, \lambda) &=& - \frac{\partial
H^{\prime}}{\partial X^a} (X^a, P_a, \lambda).  \label{exacteqns} \end{eqnarray}

\noindent These equations are exact.  We note that if the perturbation, $\Delta H$, was 
zero,
the solution to Eqs.(\ref{exacteqns}) would be that $X^{a}$ and $P_{a}$ are constants.  When
$\Delta H$ is not zero, these ``constants'' become functions of initial conditions, $(X^a_o,
P_a^o)$, and the parameter, $\lambda$.  For this reason, this method is referred to as the
``variation of constants.''

In principle, the solution of the equations of motion, Eqs.(\ref{exacteqns}), would be $2n$
functions of the form

\begin{eqnarray} X^{a} &=&X^{a}(X_{o}^{a},P_{a}^{o},\lambda ) \nonumber \\ P_{a}
&=&P_{a}(X_{o}^{a},P_{a}^{o},\lambda ), \label{sol1} \end{eqnarray}

\noindent where $X_{o}^{a}$ and $P_{a}^{o}$ are $2n$ initial values.  Substituting
Eqs.(\ref{sol1}) into the canonical transformation, Eqs.(\ref {canntrans}), yields the
solution to Hamilton's equations for $x^{a}$ and $p_{a}$ as functions of initial conditions
and $\lambda $.

Note that we have made no assumption about the value of the unperturbed Hamiltonian.  As an
example, suppose that we wish to solve a Kepler type problem, where $H_{o}$ represents a
spherically symmetric Hamiltonian and $\Delta H$ represents a perturbation.  Since the value
of $H_{o}$ has not been determined, this method applies to perturbations of bounded or
unbounded orbits.

Since the equations for $X^{a}$ and $P_{a}$, Eqs.(\ref{exacteqns}), are exact, they tend to
be as difficult to solve as Hamilton's equations in the original phase space variables,
$(x^{a},p_{a})$.  Typically, the variation of constants method is not used to find exact
solutions, but rather it leads to an approximation procedure.

The natural approximation procedure associated with the variation of constants method is to
solve Hamilton's equations, Eqs.(\ref{exacteqns}), by successive iteration.  To iterate,
one must choose an appropriate initial trajectory, $(X^a_o (\lambda), P_a^o (\lambda))$, as 
the ``zeroth iterate'' and insert
these functions of $\lambda$ into the right hand side of Eq.(\ref{exacteqns}).  In this
way, the right hand sides of Eqs.(\ref{exacteqns}) become functions of the parameter 
$\lambda
$.  The first iterate approximation is then obtained by simple integration on $\lambda$.  We 
want to emphasize
that {\it{any}} trajectory, suitably close to the exact result, can be used as the zeroth
iterate.  In practice, general knowledge of the physical situation will guide the choice of
the zeroth iterate.

One natural choice is to take the zeroth iterate as given by $X^{a} $ and $P_{a}$ as
constants, $X^{a}=X_{o}^{a}$ and $P_{a}=P_{a}^{o}$, which is the result obtained when 
$\Delta
H = 0$.  The first iterate then is given by

\begin{eqnarray} X_{1}^{a}(X_{o}^{a},P_{a}^{o},\lambda ) &=&X_{o}^{a}+\int_0^{\lambda
}d\lambda ^{\prime }\left( \frac{\partial H^{\prime }}{\partial P_{a}^{o}}
(X_{o}^{a},P_{a}^{o},\lambda ^{\prime })\right) \nonumber \\
P_{a}^{1}(X_{o}^{a},P_{a}^{o},\lambda ) &=&P_{a}^{o}-\int_0^{\lambda }d\lambda ^{\prime
}\left( \frac{\partial H^{\prime }}{\partial X_{o}^{a}} (X_{o}^{a},P_{a}^{o},\lambda 
^{\prime
})\right) .  \label{X1P1} \end{eqnarray}

As we will discuss in Section~IV, this choice is not the most appropriate initial 
trajectory
for the application to lensing.  Instead, we will take a sequence of constants, $(X^a_o,
P^o_a)$, for different ranges in $\lambda$ as the zeroth iterate.  These constants will be
selected from values predicted by the thin lens approximation.

The $n$th iterate is given in the same way:  $X^{a}$ and $P_{a}$ are replaced by
$X_{n-1}^{a}(X_{o}^{a},P_{a}^{o},\lambda ^{\prime })$ and
$P_{a}^{n-1}(X_{o}^{a},P_{a}^{o},\lambda ^{\prime })$ on the right hand side of
Eqs.(\ref{exacteqns}) after the derivative is taken.  In this way, the right hand sides of
Eqs.(\ref{exacteqns}) are always functions of $\lambda^{\prime }$ and initial values,
$(X_{o}^{a},P_{a}^{o})$, and can be integrated immediately.  The $n$th iterate approximate
solution to Hamilton's equations in the original phase space variables is obtained by
substituting $ X_{n}^{a}$ and $P_{a}^{n}$ into the canonical transformation, Eqs.(\ref
{canntrans}).

In the limit that $n\rightarrow \infty$, the iterative solution, $(X^a_n, P^n_a)$, always
converges to the exact solution for small $\lambda$ values if the initial trajectory is 
chosen as constant values.  However, even in these
converging cases, at {\it{any}} finite $n$, the iterative solution may differ greatly from
the exact solution at large $\lambda$ values.  For example, Goldstein~\cite{Goldstein}
considers a harmonic oscillator perturbation,

\[ \Delta H = \frac{1}{2} m \omega^2 x^2, \]

\noindent on a free Hamiltonian in one dimension,

\[ H_o = \frac{p^2}{2m}.  \]

\noindent Using particular initial conditions, he finds that the $2$nd iterate solution in
the original variables given by

\[ x_2 = \frac{p_o}{m\omega}\left(\omega t - \frac{\omega^3 t^3}{3!}  + \frac{\omega^5
t^5}{5!}\right), \]

\[ p_2 = p_o \left(1 - \frac{\omega^2 t^2}{2!}  + \frac{\omega^4 t^4}{4!}\right), \]

\noindent As $n\rightarrow \infty$, the limit converges to

\[ x\rightarrow \frac{p_o}{m\omega} \sin{\omega t} \quad\quad p \rightarrow p_o \cos{\omega
t}.  \]

\noindent However, at any finite $n$, the solution runs away to infinity, quickly separating
from the exact result.  This is because the finite $n$ iterative solution is simply the 
first
few terms in a Taylor series expansion of the exact result.

To develop a method which gives reasonable results at finite $n$, one may attempt to adjust
the standard iterative method using general knowledge of the features of the exact solution.
Such corrections are required in the application of the variation of constants method to
gravitational lensing when accurate results at low $n$ are desired.  These corrections, 
which
will be discussed in Section~IV.C, involve the consecutive use of a series of unperturbed
orbits.

\section{Application to Gravitational Lensing}

The iterative approach to gravitational lensing will be an approximation to the exact
approach and will remain within its overall perspective and framework \cite {FN,EFN}.  In
Section~III.A, we outline the key elements of the exact approach, while the iterative
approach will be developed in Section~III.B.  We show how to turn null geodesics into lens
equations in Section~III.C.  For clarity, we leave any corrections of the iterative method 
at
low $n$ to the discussion in the final subsection of Section~IV.

\subsection{Exact Gravitational Lensing}

In the exact approach, one considers a space-time, $(M,g^{ab})$, containing various lenses 
(travelling on specified world lines) whose properties are hidden in the metric $g^{ab}$, 
and an observer moving on a world line described by

\begin{equation} x^a = X^a_o (\tau).  \end{equation}

\noindent The past light-cone of this observer is generated by null geodesics originating at
$X^a_o(\tau)$, and may be described parametrically as

\[ x^{a}=X^{a}(\tau ,\theta ,\phi ,\lambda ) \]

\noindent or

\begin{eqnarray} t &=&{\cal T}(\tau ,\theta ,\phi ,\lambda ) \label{timeofflight} \\ x^{i}
&=&{\cal X}^{i}(\tau ,\theta ,\phi ,\lambda )\quad \quad i=1,\,2,\,3.  \label{lenseqns}
\end{eqnarray}

\noindent In these equations, the proper time, $\tau $, labels the points along the world
line of the observer, and $\lambda $ represents an affine parameter along the null geodesics
of the observer's past light-cone.  The angles, $(\theta ,\,\phi)$, represent the sphere of
null directions seen by the observer, or the observer's celestial sphere.

The time equation, Eq.(\ref{timeofflight}), describes the coordinate emission time of a 
light
ray emitted by a source at $x^{i}$ arriving at the observer at time $\tau$.  This equation,
often referred to as the ``time of arrival equation,'' is important in astrophysics, among
other reasons, because it can be used, in conjunction with observations and a lens model, to
place bounds on the Hubble constant, see for example~\cite{Impey,Chae}.

The three space-like equations, Eqs.(\ref{lenseqns}), can be considered as a generalized,
exact form of the gravitational lens equations, because they map observed directions into
source positions.  By giving a relationship between the affine parameter and some observable
distance scale -- possibly the redshift distance, $R$~\cite{FKN} -- one can express the 
generalized,
exact lens equations as

\begin{equation} x^{i} ={\cal X}^{i}(\tau ,\theta ,\phi ,R)\quad \quad i=1,\,2,\,3,
\label{exactlens} \end{equation}

\noindent which is similar the traditional form of lens equations, where the $(\tau, \theta,
\phi, R)$ are all observable quantities.  In \cite{FN,EFN}, Eq.(\ref{exactlens}) is refered
to as the exact lens equation.  This step is necessary for comparison wilowth astrophysical
observations because the affine parameter has no measurable meaning, but will not be 
required
in our discussion.

As mentioned, the general perspective taken in exact gravitational lensing is that the lens
properties are built into the form of the metric, $g^{ab}$.  This means that there can be no
reference to any background space-time or any quantity ``in the absence of the lens.'' To
find the explicit form of the lens equations and time of arrival equation, one must solve 
the
null geodesic equations associated with $g^{ab}$, giving initial conditions for past 
directed
rays on the observer's world line.

Therefore, we begin with the full geodesic equations:  four second order coupled 
differential
equations for the four base space coordinates, $x^{a}$.  These equations can alternatively 
be
viewed as a set of eight coupled first order differential equations on the cotangent bundle,
$(x^{a},p_{a})$, which are derived from a Hamiltonian.  To make them into the null geodesic
equations, initial conditions must be chosen so that the Hamiltonian vanishes.

The Hamiltonian for geodesic orbits is

\begin{equation} H = g^{ab} (x^a) p_a \, p_b, \end{equation}

\noindent with associated Hamilton's equations:

\begin{eqnarray} {\dot{x}}^{a} &=&2g^{ab}p_{b} \nonumber \\ {\dot{p}}_{a}
&=&-{g^{bc}},_{a}p_{b}\,p_{c}.  \label{grham} \end{eqnarray}

\noindent Solutions to Hamilton's equations have the form

\begin{eqnarray} x^{a} &=&{\tilde{x}}^{a}(x_o^{a},p_{a}^o ,\lambda ) \nonumber \\ p_{a}
&=&{\tilde{p}}_{a}(x_o^{a}, p_{a}^o , \lambda ), \label{sol2} \end{eqnarray}

\noindent where $x_{o}^{a}$ is an initial position representing the location of the observer
and $p_{a}^{o}$ gives the initial direction of the geodesic.  The coordinates, $x^{a}$,
represent the space-time position along the geodesic, and $p_{a}$ is the cotangent vector to
this geodesic.

Null geodesics correspond to the geodesics whose value of the Hamiltonian is zero.  Since 
the
Hamiltonian does not depend on the parameter $\lambda$, its value will be preserved along 
the
geodesics, and can be set to zero by constraining the initial conditions:

\begin{equation} H(x_o^{a},p_{a}^o)=g^{ab}(x_o^{a})p_{a}^o\,p_{a}^o = 0.  \label{initcond}
\end{equation}

\noindent Equation~(\ref{initcond}) is quadratic in $p_0^o$ and can always be solved for
$p_{0}^{o}~=~p_{0}^{o}(x_{o}^{a},p_{i}^{o})$.  Substituting this into Eqs.(\ref{sol2}) gives

\begin{equation} x^{a}={\tilde{x}}^{a}(x_o^{a},p_{{0}}^o(x_o^{a},p_{i}^o),p_{i}^o, \lambda
)=x^{a}(x_o^{a},p_{i}^o,\lambda ), \label{null} \end{equation}

\noindent which are null geodesics.

\subsection{Iterative Method for Null Geodesics}

In general, it is impossible to solve Hamilton's equations, Eqs.(\ref{grham}), in closed
form.  However, if one is able to consider the metric of physical interest, $g^{ab}$, to be
close to some metric, $g_{o}^{ab}$, for which one can solve Hamilton's equations, then the
method of variation of constants gives a way to systematically approximate the null
geodesics.

Formally, we suppose that we have two space-times, $(M,g^{ab})$ and $(M,g_{o}^{ab})$ where

\begin{equation} g^{ab}=g_{o}^{ab}+h^{ab}, \end{equation}

\noindent and $h^{ab}$ is ``small.'' For the sake of discussion, we will suppose that
$(M,g_{o}^{ab})$ represents either Minkowski space-time or a Robertson-Walker geometry, 
since
these are the usual background space-times in gravitational lensing.  We form the associated
Hamiltonians in both space-times:

\begin{equation} H_{o}(x^{a},p_{a}) =g_{o}^{ab}\,(x^{a})p_{a}\,p_{b} \label{unperturbed}
\end{equation}

\begin{equation} H(x^{a},p_{a}) =g^{ab}(x^{a})p_{a}\,p_{b}=g_{o}^{ab}\,(x^{a})p_{a}
\,p_{b}+h^{ab}\,(x^{a})p_{a}\,p_{b}.  \label{perturbed} \end{equation}

By finding a complete solution to the Hamilton-Jacobi equation in $(M,g_{o}^{ab})$,

\begin{equation} g_{o}^{ab}\frac{\partial F_{o}}{\partial x^{a}}\,\frac{\partial F_{o}}{
\partial x^{b}}+\frac{\partial F_{o}}{\partial \lambda } = 0, \label{grH-J} \end{equation}

\noindent we obtain a generating function of canonical transformations,
$F_{o}(x^{a},P_{a},\lambda )$.  Because Minkowski space-time and the Robertson-Walker
geometries are integrable geometries, we will always be able to solve Eq.(\ref{grH-J}) and
find an explicit canonical transformation in the form of Eqs.(\ref{canntrans}):

\begin{eqnarray} x^a &=& {\tilde{x}}^a (X^a, P_a, \lambda) \label{canntrans2a} \\ p_a &=&
{\tilde{p}}_a (X^a, P_a, \lambda).  \label{canntrans2b} \end{eqnarray}

Using the explicit form of the canonical transformation, Eqs.(\ref {canntrans2a},
\ref{canntrans2b}), we transform the Hamiltonian in the perturbed space,
Eq.(\ref{perturbed}), obtaining

\begin{eqnarray} H^{\prime }(X^{a},P_{a},\lambda ) &=&H(X^{a},P^{a},\lambda )+\frac{\partial
F_{o}}{\partial \lambda } \nonumber \\ &=& \left( h^{ab} 
({\tilde{x}}^{a}(X^{a},P_{a},\lambda
)) \right){~\tilde{p}} _{a}(X^{a},P_{a},\lambda )\,{~ \,}{\tilde{p}}_{b}(X^{a},P_{a}, 
\lambda
).  \label{grH'} \end{eqnarray}

\noindent Hamilton's equations of motion for $X^a$ and $P_a$ are

\begin{eqnarray} {\dot{X}}^{a}(X^{a},P_{a},\lambda ) &=& 2\left( h^{bc} \left({\tilde{x}}
^{a}(X^{a},P_{a},\lambda )\right) \right) \, {\tilde{p}}_{b}(X^{a},P_{a},\lambda ) \left(
\frac{\partial {\tilde{p}}_{c}} {\partial P_a} (X^{a},P_{a},\lambda ) \right) \nonumber \\ &
& ~ + \left( \frac{ \partial h^{bc} }{\partial P_a} ({\tilde{x}} ^{a}(X^{a},P_{a},\lambda))
\right) {\tilde{p}}_{b}(X^{a},P_{a},\lambda ) {\ \tilde{p}}_{c}(X^{a},P_{a},\lambda )
\nonumber \\ & \equiv & \Xi^a (X^a, P_a, \lambda) \label{Xdot} \end{eqnarray}

\begin{eqnarray} {\dot{P}}_{a}(X^{a},P_{a},\lambda ) &=&-\left( \frac{\partial
h^{bc}}{\partial X^a} \left({\tilde{x}} ^{a}(X^{a},P_{a},\lambda ) \right) \right) \,
{\tilde{p}} _{b}(X^{a},P_{a},\lambda\, ) \,{\tilde{p}}_{c}(X^{a},P_{a},\lambda ) \nonumber 
\\
& & ~\left( -2 h^{bc}\left ({\tilde{x}} ^{a}(X^{a},P_{a},\lambda ) \right) \right) \, {\
\tilde{p}}_{b}(X^{a},P_{a},\lambda ) \left( \frac{\partial {\tilde{p}}_{c}} { \partial X^a}
(X^{a},P_{a},\lambda ) \right) \nonumber \\ & \equiv &\Upsilon_a (X^a, P_a, \lambda).
\label{Pdot} \end{eqnarray}

We now use the iterative procedure of Section~II to solve these equations.  For the moment, 
we will choose the zeroth
iterate as $X^{a}$ and $P_{a} $ taking constant values,
$(X_{o}^{a},P_{a}^{o})$, under the integral.  The $n$th iterate will be given by

\begin{eqnarray} X_{n}^{a}(X_o^{a},P_{a}^o,\lambda ) &=&X_o^{a}+\int_0^{\lambda }d\lambda
^{\prime }~\Xi^a (X_{n-1}^{a},P_{a}^{n-1},\lambda ^{\prime }) \label{grXnth} \\
P_{a}^{n}(X_o^{a},P_{a}^o,\lambda ) &=&P_{a}^o+\int_0^{\lambda }d\lambda ^{\prime
}~{\Upsilon}_{a}(X_{n-1}^{a},P_{a}^{n-1},\lambda ^{\prime }) , \label{grPnth} \end{eqnarray}

\noindent where $X^a_{n-1}$ and $P_a^{n-1}$ are functions of initial conditions and
$\lambda^{\prime}$.

When the $n$th iterate is substituted into Eq.(\ref{canntrans2a}), the result is an
approximate solution to the geodesic equations of $(M,g^{ab})$ with initial conditions
$(X_o^{a},P_{a}^o)$:

\begin{equation} x^{a}={\tilde{x}}^{a}(X_{n}^{a}(X_o^{a},P_{a}^o,\lambda
),P_{a}^{n}(X_o^{a},P_{a}^o,\lambda ),\lambda ).  \label{sol3} \end{equation}

\noindent These approximate geodesics may not be approximately null.  In analogy to the 
exact
case, we should require that

\begin{equation} H(X_o^{a},P_{a}^o) = 0, \label{initcond2} \end{equation}

\noindent which we can solve for $P_{0}^o$ as a function of the other initial conditions,

\begin{equation} P_0 = P_{0}^o(X_o^{a},P_{i}^o).  \label{P0first} \end{equation}

\noindent Substituting $P_{0} ^o(X_o^{a},P_{i}^o)$ into Eq.(\ref{sol3}) gives approximate
null geodesics:

\begin{equation} x^{a}={\tilde{x}}^{a}(X_{n}^{a}(X_o^{a},P_{i}^o,\lambda
),P_{a}^{n}(X_o^{a},P_{i}^o,\lambda ),\lambda ) = x^{a}(X_o^{a},P_{i}^o,\lambda ).
\label{lensandtime} \end{equation}

It is important to realize that, while the iteration method introduces a ``background
space-time,'' $(M, g_o^{ab})$, the null geodesics one seeks are the null geodesics of the
physical space-time, $(M, g^{ab})$.  The spirit of exact gravitational lensing is that no
reference to the background should be made.  In the iterative approach, a background is
introduced {\it{only}} to facilitate finding the approximate geodesics of $(M, g^{ab})$ and
plays no further role.  All physical quantities are computed in the physical space-time 
using
$g^{ab}$, not $g_o^{ab}$; the null condition is fixed by solving Eq.(\ref{initcond2}) for
$P_0^o$, in which the physical metric appears.  Likewise, angles are to be computed from the
inner product of two vectors which lie in the tangent space of $(M, g^{ab})$.  Hence, the
iterative approach remains within the spirit of the exact approach because it does not refer
in any physical way to a background space-time.

The $n$th iterate, approximate null geodesics can thus be used as the basis for an
approximate gravitational lensing theory under the spirit of the exact approach.  For these
purposes, the $X_{o}^{a}$ in Eq.(\ref{lensandtime}) must be chosen as the space-time 
position
representing the location of an observer, and the $P_{i}^{o}$ are used to parametrize the
directions which an observer can see.  Sources are then located at $x^{a}$, along the past
directed null geodesics described by Eq.(\ref{lensandtime}).  In the next subsection, we
describe the procedure for turning (either an exact or an approximate) solution to the null
geodesic equations into the form of a time of arrival equation and lens equations given by
Eqs.(\ref{timeofflight}, \ref{lenseqns}).

\subsection{Exact and Approximate Lens Equations}

In this subsection, we describe how either an exact or an approximate solution to the null
geodesic equations in an arbitrary space-time gives rise to gravitational lens equations.
The null geodesics, exactly or approximately, are described by

\begin{eqnarray} x^a &=& x^a (x_o^a, p_i^o, \lambda) \label{solx1} \\ p_a &=& p_a (x^a_o,
p_i^o, \lambda), \label{solp1} \end{eqnarray}

\noindent where the $x_{o}^{a}$ is an initial space-time location, the three $p_{i}^{o}$
describe the spatial direction of the null geodesic at that point, and $\lambda $ is an
affine parameter along the null geodesic, where $\lambda =0$ corresponds to the initial
position.  The null tangent vectors to these geodesics, $p^{a}=g^{ab}p_{b}={\dot{x}}^{a}$,
are taken as past directed.  The scaling of $\lambda$ has not yet been chosen.

To turn Eqs.(\ref{solx1}) into a time of arrival equation and lens equations in the form of
Eqs.(\ref{timeofflight}, \ref{lenseqns}), we introduce an observer by taking a time-like
curve parametrized by the proper time, $\tau $,

\begin{equation} x^a = X^a_o (\tau), \end{equation}

\noindent as a one parameter family of initial positions in Eqs.(\ref{solx1}), or
$x_{o}^{a}=X_{o}^{a}(\tau )$, for the past directed null geodesics.

The tangent vector to this world line,

\[ v^a = \frac{ d X^a_o (\tau)}{ d \tau} \]

\noindent can be used to scale the $p_{i}^{o}$ (and consequently $\lambda$).  Because the
magnitude of the $p_{i}^{o}$ is irrelevant, we can fix its length by requiring that

\begin{equation} v^{a}p_{a}^{o}=-1.  \label{norm} \end{equation}

\noindent This implies that, at any point along the observer's world-line, the three
$p_{i}^{o}$ are functions of only two parameters.  These two parameters can be taken as two
independent ``observation angles,'' $(\theta, \phi)$.  Thus, the null cotangent vector at
$X_{o}^{a}(\tau )$ can be expressed as

\begin{equation} p_{a}^{o}=p_{a}^{o}(\theta ,\phi ).  \label{pa} \end{equation}

By replacing $x^a_o$ with $X^a_o (\tau)$ and $p_i^o$ with Eq.(\ref{pa}), in 
Eq.(\ref{solx1}),
one obtains a time of arrival equation and lens equations in the form of
Eqs.(\ref{timeofflight}, \ref{lenseqns}):

\begin{eqnarray} t &=&x^{{0}}(X_{o}^{a}(\tau ),p_{i}^{o}(\theta ,\phi ),\lambda ) \nonumber
\\ &\equiv &{\cal T}(\tau ,\theta ,\phi ,\lambda ) \nonumber \\ x^{i} 
&=&x^{i}(X_{o}^{a}(\tau
),p_{i}^{o}(\theta ,\phi ),\lambda ) \nonumber \\ &\equiv &{\cal X}^{i}(\tau ,\theta ,\phi
,\lambda )\quad i=1,\,2,\,3.  \label{lenseqns2} \end{eqnarray}

As an example, we consider an observer moving along a world line given by $X_{o}^{a}(\tau
)=(\tau ,0,0,+z_{o})$ in a linearized Schwarzschild space-time.  From Eq.(\ref{norm}), we 
see
that $p_{0}^{o}(x_{o}^{a},\,p_{i}^{o})=-1$ for a past directed null geodesic.  This
normalization can be obtained in the previous section by dividing all components of $P_a$ by
the constant value of $P_0$ given in Eq.(\ref{P0first}).

Since $p_{a}^{o}$ is null, with $p_{0}^{o}=-1,$ its spatial part will have the length:

\begin{equation} |p_i^o| = \sqrt{ - g^{ij}\, p_i^o \, p_j^o }= \sqrt{1+ \frac{2m}{|z_o|}}.
\end{equation}

\noindent Thus, we can write the spatial part of the null covector at the observer as

\[ p_i^o = \sqrt{1+ \frac{2m}{|z_o|}} u_i (\theta, \phi), \]

\noindent where $u_i (\theta, \phi)$ is a unit vector parametrized by two observation 
angles,
$(\theta, \phi)$.

From the spherical symmetry of Schwarzschild space-time, the lens, source, and observer can
be taken as lying in the same spatial plane, and any ray with an initial direction in this
plane will remain in the plane.  Fixing this plane as the $\hat{x}$-$\hat{z}$ plane fixes 
one
of the two observation angles, say $\phi =0$.  An observer on the $+\hat{z}$ axis will
observe the lens located at the origin in the direction of the unit spatial vector,

\[ n^{i}=\frac{- x^i_o}{\sqrt{-g_{ij}\,x^i_o \, x^j_o}} = \frac{-\delta^{i \, 3}}{\sqrt{1 -
\frac{2m}{|z_o|}}}.  \]

\noindent In terms of $u_{i}$ and $n^{i}$, the remaining observation angle, $\theta$, is
defined by

\begin{equation} \cos {\theta }= u_i\,n^i.  \label{obsangle} \end{equation}

\noindent Since $n^i$ is proportional to $\delta^{i\,3}$, we have

\[ u_z = -\sqrt{1 - \frac{2m}{|z_{o}|}} \cos{\theta}.  \]

\noindent Using $\sqrt{-g^{ij}\, u_i \, u_j} = |u| = 1$, $u_y = 0$, and the linearized 
metric
given in Section~IV.B, one finds the final component on the momentum,

\[ u_x = \sqrt{1 - \left(1-\frac{2m}{|z_{o}|}\right)^2 \cos^2{\theta}}.  \]

We note that the observation angle is determined relative to the physical metric of the
space-time, $g^{ab}$.  This is consistent with the notion that the iterative and exact
methods should not make reference to a background space-time.

\section{The Schwarzschild Lens}

To demonstrate the iterative method, we apply this procedure to the case of the 
Schwarzschild
lens, considering the Schwarzschild metric as a perturbation of Minkowski space-time.
Throughout this section we will work in $(t,x,y,z)$ coordinates which, while awkward for the
Schwarzschild metric, facilitate the iterative procedure and comparisons with the thin lens
approach.  In the iterative calculation, we work with the linearized version of the
Schwarzschild metric, since there are no significant differences for the large impact
parameters usually discussed.

The section will be divided into three subsections.  First, we outline the thin lens
approximation and its application to the Schwarzschild lens.  The second subsection develops
the application of the iterative method to the Schwarzschild lens.  Finally, we discuss
the choice of the zeroth iterate when one wishes to use low $n$ values. This subsection
makes use of results from both of the first two subsections.

\subsection{The Thin Lens Approximation}

Here, we briefly outline the standard thin lens approximation.  Most theoretical details
about the thin lens approximation can be found in the book by Falco, Schneider and
Ehlers~\cite{Ehlers}.

\begin{figure}[tbp] \begin{center} \scalebox{1.0}{\includegraphics{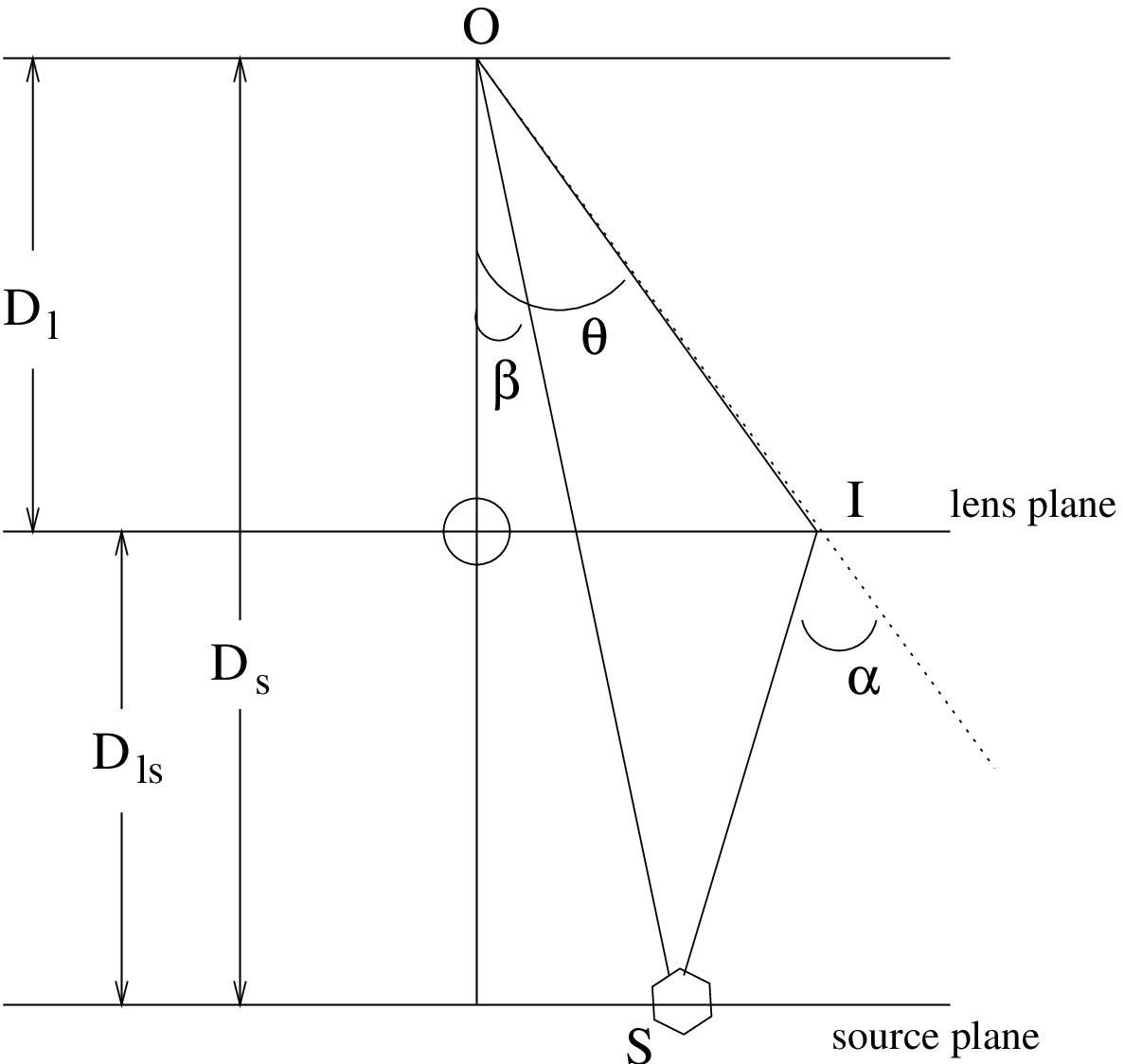}} \end{center}
\caption{The general picture used in the thin lens approximation.  A past directed geodesic
emitted at the observer O travels rectilinearly to I, is bent by the bending angle, 
$\alpha$,
and then travels to a source at S.  Distances $D_{l}$, $D_{s}$, and $D_{ls}$ separate the
lens and observer, source and observer, and lens and source, while $\beta $ locates the
source and $\theta $ is an observation angle.}  \label{thin} \end{figure}

The key assumption in the thin lens approximation is that the null geodesics in the
space-time of interest travel along the null geodesic path of the background space except at
isolated points near the lens where sharp bending occurs.  For a Schwarzschild lens in a
Minkowski background, the geodesic travels in rectilinear motion from the observer, O, to a
point in the lens plane, I, and then from I to the source location at O, as is shown in
Fig.~\ref{thin}.

The effect of the lens is twofold.  First, the trajectory is instantaneously bent at I by a
bending angle computed as the angle between the two asymptotic paths of the true null
geodesic.  For a Schwarzschild lens, the bending angle is simply $\alpha = 4\,G\, m\,/(c^2 
\,
r_o)$, where $r_o$ is the radius of closest approach, $m$ is the mass, and the physical
factors $G$, and $c$ have been restored.  Second, the lens affects the time of arrival by
introducing a gravitational time delay relative to the time required to traverse the same
path in the Minkowski background.

The lens equation (of the thin lens approximation) is the map from the space of observed
light-ray directions to the source positions (on the source plane).  Often the map is not
invertible, as there could be multiple images for a given point source.  In practice, one
develops a model for the lens, usually in terms of a potential for the mass distribution, 
and
from this model determines the amount of bending a null geodesic will encounter along its
trajectory from past null infinity to future null infinity.  This bending angle, $\alpha$, 
in
Fig.~\ref{thin}, is amount of bending of the ray at the lens plane.  An angle $\beta$ 
locates
the true source location in the background space-time.

For a single Schwarzschild lens, one determines from Euclidean geometry that the lens
equation is

\begin{equation} \beta =\theta -\frac{4\, G\, m\,D_{ls}}{c^2 \, D_{s}\,D_{l}\,\theta },
\label{thinlens} \end{equation}

\noindent where $\theta$ is the image observation angle, and $D_{l}$, $D_{s}$, and $D_{ls}$
represent the distances between the lens and observer, the source and observer, and the lens
and source, respectively.  Generically, there will be two values of $\theta$ for a given
value of $\beta$, so that there will be two distinct paths joining the source and observer,
with one passing on each side of the lens.

In the thin lens approximation, the time of arrival is determined by the coordinate time
which elapses along the path taken in the thin lens approximation.  Although the path, up to
a sharp bending in the lens plane, is determined by the background space-time, the time of
arrival is computed relative to the (usually linearized) physical metric including the lens.
The time of arrival for a ray emitted from the source at time $t_{s}$ and passing close to a
Schwarzschild lens is \cite{Ehlers}

\begin{equation} t = t_s + \frac{1}{c}\int~ \left( 1 + \frac{2 \,G\, m}{ c^2 \, r(l)}\right)
\, dl, \label{tltime} \end{equation}

\noindent where $r(l)$ represents the Euclidean distance from the lens to a point along the
trajectory.  This distance is expressed as a function of the Euclidean distance along the
trajectory, $l$, and the integral is taken along the rectilinear path from O to I to S.  A
time delay is the difference between the two times computed by Eq.(\ref{tltime}) along the
two paths determined by Eq.(\ref{thinlens}).

\subsection{Iterated Schwarzschild Null Geodesics}

This subsection develops the iterative approach in the case of the single Schwarzschild lens
along the lines of Section~III.B.  In this subsection, the zeroth iterate will be chosen as 
the solution to Hamilton's equations in the 
unperturbed metric, or $(X^a_o, P^o_a) = $ constant. A better choice of the zeroth iterate 
is proposed in the next
subsection. We begin with the linearized Schwarzschild metric, which is given by

\begin{equation} g^{ab} (x^a) = \left( \begin{array}{cccc} 1 + \frac{2m}{r} & 0 & 0 & 0 \\ 0
& -1 + \frac{ 2 m x^2}{r^3} & \frac{ 2m x y}{r^3} & \frac{2m x z}{r^3} \\ 0 & \frac{ 2 m x
y}{r^3} & -1 +\frac{2m y^2}{r^3} & \frac{ 2m y z}{r^3} \\ 0 & \frac{ 2 m x z }{r^3} & \frac{
2m y z}{r^3} & -1 + \frac{2m z^2}{r^3} \end{array} \right) \label{metric} \end{equation}

\noindent in $(t, x, y, z)$ coordinates.  Here, the signature of the metric is $(+, -, -,
-)$, and $r = \sqrt{x^2 + y^2 + z^2}$.  The Hamiltonians of Minkowski space-time and the
linearized Schwarzschild space-time are

\begin{eqnarray} H_{o}(x^{a},p_{a}) &=&\eta ^{ab}p_{a}\,p_{b}=(p_{{0} })^{2}-
\sum_{i=1}^{3}(p_{i})^{2} \label{MinkH} \\ H(x^{a},p_{a})
&=&g^{ab}(x^{a})p_{a}\,p_{b}=H_{o}(x^{a},p_{a})+h^{ab}(x^{a})p_{a}\,p_{b}, \label{SchwarzH}
\end{eqnarray}

\noindent and $h^{ab}$ is

\begin{eqnarray} h^{{0} a} &=& \frac{2 m \delta^{{0} a} }{r}, \nonumber \\ h^{ij} &=& 
\frac{2
m x^i x^j}{r^3}.  \label{hab} \end{eqnarray}

The complete solution to Hamilton-Jacobi equation of $H_{o}(x^{a},p_{a})$ is

\begin{equation} F_{o}(x^{a},P_{a},\lambda )=x^{a}P_{a}-\eta ^{ab}P_{a}\,P_{b}\lambda,
\label{MinkF} \end{equation}

\noindent and the canonical coordinate transformation associated with Eq.(\ref{MinkF}) is
given by

\begin{eqnarray} x^a &=& X^a + 2 \eta^{ab} P_b \lambda \equiv {\tilde {x}}^a(X^a, P_a,
\lambda) \label{xcanntrans} \\ p_a &=& P_a.  \label{pcanntrans} \end{eqnarray}

\noindent Applying this canonical transformation to the Hamiltonian $H(x^a, p_a)$ yields

\begin{equation} H^{\prime }(X^{a},P_{a},\lambda )=h^{ab}\left( {\tilde x}^{a}(X^{a},P_{a},
\lambda ) \right)\,P_{a}\,P_{b}, \end{equation}

\noindent and Hamilton's equations of motion for $X^a$ and $P_a$ are

\begin{eqnarray} {\dot{X}}^{a}(X^{a},P_{a},\lambda ) &=&2h^{ab} \left( {\tilde x}^{a}
(X^{a},P_{a}, \lambda ) \right) P_{b} + \frac{\partial h^{bc} \left( {\tilde
x}^{a}(X^{a},P_{a}, \lambda ) \right) }{\partial P_{a}} P_{b}\,P_{c} \nonumber \\ & \equiv &
\Xi^{a}(X^{a},P_{a},\lambda ) \nonumber \\ {\dot{P}}_{a}(X^{a},P_{a},\lambda )
&=&-\frac{\partial h^{bc} ( {\tilde x} ^{a}(X^{a},P_{a},\lambda ))}{\partial
X^{a}}P_{b}\,P_{c}\equiv \Upsilon_{a}(X^{a},P_{a},\lambda ).  \label{SchwarzHeqn}
\end{eqnarray}

We solve Hamilton's equations by the iterative procedure, as in Section~III.  The zeroth
iterate will be constant values, $X^{a}=X_o^{a}=x_o^{a}$ and $P_{a}=P_{a}^o=p_{a}^o$.  The
$n$th iterate is given by

\begin{eqnarray} X_{n}^{a}(X_o^{a},P_{a}^o,\lambda ) &=&X_o^{a}+\int_0^\lambda
\,d\lambda^\prime~\Xi^{a} (X_{n-1}^{a},P_{a}^{n-1}, \lambda^\prime) \nonumber \\
P_{a}^{n}(X_o^{a},P_{a}^o,\lambda ) &=&P_{a}^o+\int_0^{\lambda} \,d\lambda_{~}^{\prime
}~\Upsilon_{a}(X_{n-1}^{a},P_{a}^{n-1}, \lambda^\prime ).  \label{Schwarzsol} \end{eqnarray}

\noindent The approximate geodesics given by Eq.(\ref{Schwarzsol}) are not necessarily null;
we must require that $H(X_{o}^{a},P_{a}^{o})=0$.  Using the metric, Eq.(\ref{metric}), and
the Hamiltonian, Eq.(\ref{SchwarzH}), yields

\begin{equation} P_{{0}}^{o}=\left( \frac{\sum_{i=1}^{3}(P_{i}^{o})^{2}~+~\frac{2m}{
r_{o}^{3}} (X_{o}^{i}P_{i}^{o})^{2}}{1+\frac{2m}{r_{o}}}\right) ^{1/2} \label{P0}
\end{equation}

\noindent where $r_{o}=\sqrt{\delta _{ij}\,X_{o}^{i}\,X_{o}^{j}}$.  When
Eqs.(\ref{Schwarzsol}) and Eq.(\ref{P0}) are substituted into the canonical transformation,
Eqs.(\ref{xcanntrans}), the $n$th iterate, initially null geodesics are obtained.

Finally, we note that when the initial conditions satisfy the null condition, Eq.(\ref 
{P0}),
the resulting trajectories are approximately null in $(M,g^{ab})$.  Geodesics in the
unperturbed background with these same initial conditions will be space-like.  Hence, a cone
which is null in $(M,g_{o}^{ab}) $ will be fully inside the light cone of $(M,g^{ab})$ due 
to
the converging effect of the positive mass in the Schwarzschild lens.

\subsection{Choice of the zeroth iterate for low $n$}

As mentioned at the end of Section~II, the iterative method can fail to produce accurate
results at low $n$ and large $\lambda$ values regardless of whether the method converges in
general.  There are two ways in which such a failure can be anticipated in the application 
to
lensing.

First,  when the zeroth order trajectory is chosen as one set of constants, $X^a =
X^a_o$ and $P_a = P_a^o$, the trajectory is a ``straight line,'' corresponding to a
solution to the equations of motion of the unperturbed Hamiltonian.  In the Schwarzschild
case, the first order trajectory is a ``curve'' which bends in the vicinity of the lens.  
These
two trajectories become increasingly separated as $\lambda$ grows.  For $n=1$, the 
integrands
in Eqs.(\ref{Schwarzsol}) are evaluated along the zeroth order trajectory, and the ``force''
acting on the ``photon'' becomes increasingly incorrect as $\lambda$ grows.  Hence, when the
zeroth order trajectory is far from the first iterate trajectory, one may anticipate poor
results.

It turns out that for a single Schwarzschild lens, this particular source of error is
reasonably insignificant because the value of the ``force'' is decreasing faster than
$1/\lambda$ as $\lambda$ grows.  However, in a situation where there are multiple lenses
lying in more than one lens plane, the error becomes very significant, and the force vector
may point in the wrong direction altogether.

A second way to anticipate troubles is to note that while the true Hamiltonian is conserved
under a flow along the actual geodesic, it will not be conserved under an approximate flow.
If one finds the {\it{exact}} solution for $(X^a, P^a)$, and sets the initial conditions as
in Eq.(\ref{lensandtime}), the value of the perturbed Hamiltonian, Eq.(\ref{perturbed}), 
will
remain zero along the trajectory.  However, at any finite $n$, the value of

\begin{equation} H(X^a_o, P_i^o, \lambda) = H(\,x_n^a(X_o^{a},P_{i}^o,\lambda), \,
p^n_a(X_o^{a},P_{i}^o,\lambda) \, ) \end{equation}

\noindent will drift away from zero because the $n$th iterate solution is not truly a null
solution to the geodesic equations.  Hence, at large $\lambda$, the $n$th iterate solution
may not closely approximate a null geodesic.

The basic reason for these problems is that, so far, we have chosen zeroth iterate in a
non-physical way, as a straight line path corresponding a single solution to the equations 
of
motion for geodesics in the unphysical space-time, $(M, g^{ab}_o)$.  This solution separates
from the first iterate shortly after passing the lens plane.

The simplest choice of the zeroth iterate is the solution of the
equations of motion of the unperturbed Hamiltonian, but on the other hand, the zeroth 
iterate should remain close to the first iterate trajectory.  As we have discussed, the thin 
lens path is given by two connected solutions of the equations of motion in $(M, g^{ab}_o)$ 
and remains close to the exact solution.

Therefore, we may avoid these problems by {\it{choosing}} the thin lens path as the zeroth
iterate. This is accomplished by choosing different sets of constant values, $(X^a_o,
P_a^o)$, in different $\lambda$ ranges as described below.  Making this choice insures the 
first iterate
trajectory will always remain close to the zeroth iterate regardless of the number of
lenses.

From a kinematical standpoint, the entire thin lens trajectory is determined by the choice 
of
$X^i_o$ and $P_i^o$.  These values determine the straight line path from the observer at O 
to
I in Fig.~\ref{thin}.  They also determine a new set of six parameters, $(X^i_{lp},
P_i^{lp})$ which represent the location of the intersection of the original straight line
path and the lens plane, $(lp)$, and the new momenta determining the straight line path
joining I and S.

As Schwarzschild space-time is static, the $X^0 = T$ and $P_0$ equations separate from the
spatial equations.  Due to spherical symmetry, we can assume that the spatial path of the
geodesic lies in the $\hat x$-$\hat z$ plane.  In this case, the thin lens trajectory is
determined by

\begin{eqnarray} x &=& X^1_o - 2 \,P_1^o \lambda \quad\quad 0<\lambda<\lambda_l \label{xtl1}
\\ z &=& X^3_o - 2\, P_3^o \lambda \quad\quad 0<\lambda<\lambda_l \label{ztl1} 
\end{eqnarray}

\noindent and

\begin{eqnarray} x &=& X^1_{lp} - 2\,P_1^{lp} \lambda \quad\quad
\lambda_l<\lambda<\lambda_{source} \label{xtl2} \\ z &=& X^3_{lp} - 2\,P_3^{lp} \lambda
\quad\quad \lambda_l<\lambda<\lambda_{source}.  \label{ztl2} \end{eqnarray}

\noindent If the initial conditions represent an observer on the $\hat z$ axis at $X^1_o =
0$, and $X^3_o = z_o$, the lens plane will be $z = 0$, and we have

\begin{eqnarray} \lambda_l &=& \frac{z_o}{2\, P_3^o} \nonumber \\ X^1_{lp} &=& \frac{- P_1^o
\, z_o}{P_3^o} \nonumber \\ x^3_{lp} &=& 0.  \end{eqnarray}

The scaling for $\lambda$ and the spatial part of the $p_a$ is determined by

\begin{equation} v^a p_a = - 1, \label{thinlength} \end{equation}

\noindent as in Section~III.C.  Then, by choosing the new $P_i^{lp}$ such at the angle
between $P^{lp}_i$ and the original $P_i^o$ at $X^i_{lp}$ is the bending angle, $\alpha =
4\,G\,m / (c^2 \, r_o)$,

\begin{equation} \cos{\alpha} = \frac{g^{ij}_o\, P_i^{lp} \, P_j^o}{\sqrt{(g^{ij}_o\,
P_i^{lp} \, P_j^{lp})\,\,(g_{ij}^o\, P_i^o \, P_j^o)}}, \label{thinangle} \end{equation}

\noindent both the $P_i^{lp}$ are determined.

In summary, we may always pick the initial conditions as $(X^1_o = 0, X^3_o = z_o, P_1^o =
p_x^o, P_3^o = p_z^o)$.  This fixes the value of $\lambda_l$ reaching the lens plane and
$X^1_{lp}$.  Eq.(\ref{thinlength}) and Eq.(\ref{thinangle}) determine the $P_i^{lp}$ in 
terms
of $(z_o, p_x^o, p_z^o)$.  Hence, by freely choosing initial values at the observer, the
entire zeroth iterate path is determined.

The first iterate solution will then be represented by the integrals, 
Eqs.(\ref{Schwarzsol}),
where for $0<\lambda<\lambda_l$ the constant values are fixed at the initial point, and for
$\lambda_l<\lambda<\lambda_{source}$, the values are reset.  The initial $P_0^o$ is fixed as
in Eq.(\ref{P0}) at the initial point along the first leg, and then reset at $X^i_{lp}$ with
the new values.  This insures that the value of the Hamiltonian, $H = g^{ab}\,p_a \,p_b$ 
will
remain close to zero along the trajectory.

\begin{figure}[tbp] \begin{center} \scalebox{1.0}{\includegraphics{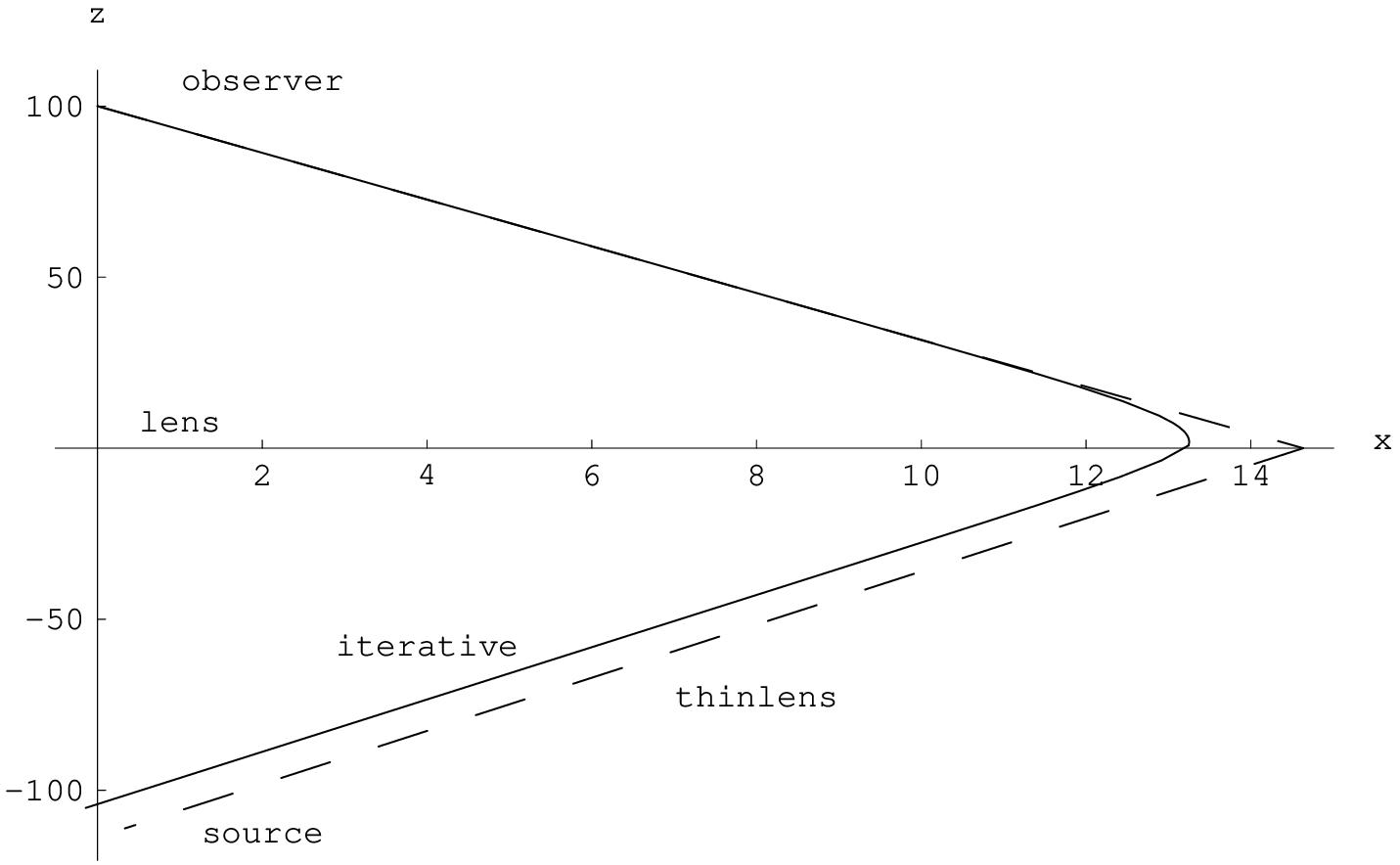}} \end{center}
\caption{The spatial path of the thin lens and first iterate approximate null geodesic when
the first iterate takes the thin lens approximation as its zeroth iterate.}  \label{path}
\end{figure}

As an example, in Fig.~\ref{path}, we plot a first iterate trajectory where the initial
conditions are taken from the thin lens path as mentioned in this section.  This figure also
shows a thin lens trajectory with similar initial conditions in dashed lines.  The mass is
taken to be $1.0$ in geometrical units and the range in $\hat z$ is $100$ units.  With these
parameter choices, the differences between the first iterate and thin lens are greatly
enhanced compared to what one would see for a more physical choice of distance scale.

We see that the first iterate trajectory is a smooth path, while the thin lens path has a
kink at the lens plane.  This demonstrates one fundamental difference between the thin lens
and iterative approaches, namely, that the iterate approach represents a continuous bending
process while the thin lens approach reduces the action of the gravitational force to a
single lens plane.  We also note that there can be a significant difference in the distance
of closest approach between the two methods, which leads to a difference in the bending 
angle
and possible location of a source.

The $n$th iterate time elapsed along the trajectory is formally given by

\begin{equation} t_n = t_o + \int \, d\lambda \,2 \,P_0 \left( 1 + \frac{2 m}{r_{n-1}}
\right), \label{nthtime} \end{equation}

\noindent from Eq.(\ref{Schwarzsol}), the form of the perturbation, $h^{ab}$, and the
canonical transformation.  Here, $r_{n-1}$ is defined as the radial coordinate distance from
the origin to the point along path of the $n-1$ iterate.  It is not difficult to compute the
second iterate for the time by first computing the first iterate trajectories above, and
using these in the right hand side of Eq.(\ref{nthtime}).  In this case, the integration 
over
$\lambda$ will also contain two segments, corresponding to the two segments above.

\section{Discussion}

Gravitational lensing is a very active field of observational research and is becoming an
increasingly important tool for probing the structure of lenses and the cosmological
parameters.  Each day, new observations and analysis expand the list of lensing candidates,
providing a rich field of study.

While the thin lens approximation has proven remarkably accurate when applied to the current
observational data, it relies heavily on the weak field and small angle approximations.  
With
the ever expanding lists of candidates and improving observational techniques, it is not
difficult to imagine that the limits of the thin lens approximation may be approached in the
future.  Virbhadra and Ellis attempt to improve upon the thin lens approximation by removing
the small angle approximations~\cite{Ellis}, and other methods may try to ``fix up'' the 
thin
lens approach from within its overall framework.  An another approach was developed by Ted
Pyne and Mark Birkinshaw~\cite{pyne1,pyne2}, where the null geodesic equations are solved
pertubatively.  One may also numerically solve the geodesic equations, as in the ``direct
integration method'' of Tomita, et al~\cite{tom1,tom2}.

The iterative method provides a different approach.  Although it works within the framework
of exact gravitational lensing, the iterative approach is reasonably easy to apply to a wide
range of lensing scenarios.  In addition, when the thin lens trajectory is taken as the
zeroth order path, the iterative method represents possible improvement upon standard 
lensing
equations, with additional accuracy obtained by increasing the level of iteration.

We are preparing a close study of the iterative and thin lens approaches in a wide variety 
of
space-time settings, including multiple lenses in multiple lens planes.  Of particular
interest are the time delays in the multiple lens plane configurations.  It is entirely
conceivable that the thin lens approximation may be accurate beyond any observational 
limits,
and no noticeable differences may be found with the iterative approach.  However, we take
courage in the fact that the history of gravitational lensing is filled with false 
prophecies
of the unlikely nature of detection.

\begin{center} {\bf{Acknowledgments}} \end{center}

\noindent The authors would like to thank Al Janis, Simonetta Frittelli, Jurgen Ehlers, and
J\"org Fraudiener for their helpful advice and suggestions.  Alejandro Perez would like to 
thank
FUNDACION~YPF.  This work was supported under grants Phy~97-22049 and Phy~92-05109.

\begin{thebibliography}{15}

\bibitem{FN} S.~Frittelli and E.~T.~Newman, Phys.  Rev.  D {\bf{59}}, 124001, (1999).

\bibitem{EFN} J.~Ehlers, S.~Frittelli, and E.~T.~Newman, {\em Gravitational Lensing From a
Space-Time Perspective}, to appear in the Festchrift for John~Stachel, Ed.  J.~Renn, Kluwer
Academic Publishers, (2000)

\bibitem{Ehlers} P.~Schneider, J.~Ehlers, and E.~E.~Falco, {\em Gravitational Lenses},
(Springer-Verlag, New York, Berlin, Heidelberg, 1992).

\bibitem{Goldstein} H.~Goldstein, {\em Classical Mechanics}, (Addison-Wesley, Reading,
Massachusetts, 1980).

\bibitem{Corben} H.~C.~Corben and P.~Stehle, {\em Classical Mechanics}, (Dover, New York,
1977.)

\bibitem{Impey} C.~D.~Impey, et al, ApJ, {\bf{509}}, 551, (1998).

\bibitem{Chae} K.~H.~Chae, ApJ, (1999).  (in press, astro-ph/9906179)

\bibitem{FKN} S.~Frittelli, T.~.P.~Kling, and E.~T.~Newman, to be published in Phys. Rev. D, 
(2000).

\bibitem{Ellis} K.~S.~Virbhadra and G.~F.~R.~Ellis, astro-ph/9904193.

\bibitem{pyne1} T.~Pyne and M.~Birkinshaw, ApJ, {\bf{415}}, 459, (1993).

\bibitem{pyne2} T.~Pyne and M.~Birkinshaw, ApJ, {\bf{458}}, 46, (1996).

\bibitem{tom1} K.~Tomita, P.~Premadi, and T.T~Nakamura, Prog.  Theor.  Phys.  Supplement No.
133, 85, (1999).

\bibitem{tom2} K.~Tomita, H.~Asada, and T.  Hamana, Prog.  Theor.  Phys.  Supplement No.
133, 155, (1999).

\end{thebibliography}
\end{document}